\documentclass[%
 reprint,
superscriptaddress,
 amsmath,amssymb,
 aps,
floatfix,
]{revtex4-2}

\usepackage{graphicx}
\usepackage{dcolumn}
\usepackage{bm}
\usepackage[mathlines]{lineno}

\begin{document}


\title{Measurement of the diffuse astrophysical neutrino flux over six seasons using cascade events from the Baikal-GVD expanding telescope}

%
%


\author {V.A.~Allakhverdyan}
\affiliation{ Joint Institute for Nuclear Research, Dubna, 141980  Russia}

\author{A.D.~Avrorin}
\affiliation{ Institute for Nuclear Research of the Russian Academy of Sciences, 
Moscow, 117312 Russia}

\author{A.V.~Avrorin}
\affiliation{ Institute for Nuclear Research of the Russian Academy of Sciences, 
Moscow, 117312 Russia}

\author{V.~M.~ Aynutdinov}
\affiliation{ Institute for Nuclear Research of the Russian Academy of Sciences, 
Moscow, 117312 Russia}

\author{Z.~Barda\v{c}ov\'{a}}
\affiliation{ Comenius University, Bratislava, 81499 Slovakia}
\affiliation{Czech Technical University, Institute of Experimental and Applied Physics, CZ-11000 Prague, Czech Republic}

\author{I.A.~Belolaptikov}
\affiliation{ Joint Institute for Nuclear Research, Dubna, 141980  Russia}

\author{E.~A.~ Bondarev}
\affiliation{ Institute for Nuclear Research of the Russian Academy of Sciences, 
Moscow, 117312 Russia}

\author{I.V.~Borina}
\affiliation{ Joint Institute for Nuclear Research, Dubna, 141980  Russia}

\author{N.M.~Budnev}
\affiliation{ Irkutsk State University, Irkutsk, 664003 Russia}

\author{V.A.~Chadymov}
\affiliation{ Independed researcher}

\author{A.S.~Chepurnov}
\affiliation{ Skobeltsyn Research Institute of Nuclear Physics, Moscow State University, Moscow, 119991 Russia}

\author{V.Y.~Dik}
\affiliation{ Joint Institute for Nuclear Research, Dubna, 141980  Russia}
\affiliation{Institute of Nuclear Physics ME RK, Almaty, 050032 Kazakhstan}

\author{A.N.~Dmitrieva}
\affiliation{National Research Nuclear University MEPHI, Moscow, Russia, 115409}

\author{\fbox{G.V.~Domogatsky}}
\affiliation{ Institute for Nuclear Research of the Russian Academy of Sciences, 
Moscow, 117312 Russia}

\author{A.A.~Doroshenko}
\affiliation{ Institute for Nuclear Research of the Russian Academy of Sciences, 
Moscow, 117312 Russia}

\author{R.~Dvornick\'{y}}
\affiliation{ Comenius University, Bratislava, 81499 Slovakia}
\affiliation{Czech Technical University, Institute of Experimental and Applied Physics, CZ-11000 Prague, Czech Republic}

\author{A.N.~Dyachok}
\affiliation{ Irkutsk State University, Irkutsk, 664003 Russia}

\author{Zh.-A.M.~Dzhilkibaev}%
 \email{djilkib@yandex.ru}
  \affiliation{ Institute for Nuclear Research of the Russian Academy of Sciences, 
Moscow, 117312 Russia}

\author{E.~Eckerov\'{a}}
\affiliation{ Comenius University, Bratislava, 81499 Slovakia}
\affiliation{Czech Technical University, Institute of Experimental and Applied Physics, CZ-11000 Prague, Czech Republic}

\author{T.V.~Elzhov}
\affiliation{ Joint Institute for Nuclear Research, Dubna, 141980  Russia}

\author{V.N.~Fomin}
\affiliation{ Independed researcher}

\author{A.R.~Gafarov}
\affiliation{ Irkutsk State University, Irkutsk, 664003 Russia}

\author{K.V.~Golubkov}
\affiliation{ Institute for Nuclear Research of the Russian Academy of Sciences, 
Moscow, 117312 Russia}

\author{T.I.~Gress}
\affiliation{ Irkutsk State University, Irkutsk, 664003 Russia}

\author{K.G.~Kebkal}
\affiliation{ LATENA, St. Petersburg, 199106  Russia}

\author{V.K.~Kebkal}
\affiliation{ LATENA, St. Petersburg, 199106 Russia}

\author{I.V.~Kharuk}
\affiliation{ Institute for Nuclear Research of the Russian Academy of Sciences, 
Moscow, 117312 Russia}

\author{S.S.~Khokhlov}
\affiliation{National Research Nuclear University MEPHI, Moscow, Russia, 115409}

\author{E.V.~Khramov}
\affiliation{ Joint Institute for Nuclear Research, Dubna, 141980  Russia}

\author{M.M.~Kolbin}
\affiliation{ Joint Institute for Nuclear Research, Dubna, 141980  Russia}

\author{S.O.~Koligaev}
\affiliation{ INFRAD, Dubna, 141981  Russia}

\author{K.V.~Konischev}
\affiliation{ Joint Institute for Nuclear Research, Dubna, 141980  Russia}

\author{A.V.~Korobchenko}
\affiliation{ Joint Institute for Nuclear Research, Dubna, 141980  Russia}

\author{A.P.~Koshechkin}
\affiliation{ Institute for Nuclear Research of the Russian Academy of Sciences, 
Moscow, 117312 Russia}

\author{V.A.~Kozhin}
\affiliation{ Skobeltsyn Research Institute of Nuclear Physics, Moscow State University, Moscow, 119991 Russia}

\author{M.V.~Kruglov}
\affiliation{ Joint Institute for Nuclear Research, Dubna, 141980  Russia}

\author{V.F.~Kulepov}
\affiliation{ Nizhny Novgorod State Technical University, Nizhny Novgorod, 603950 Russia}

\author{A.A.~Kulikov}
\affiliation{ Irkutsk State University, Irkutsk, 664003 Russia}

\author{Y.E.~Lemeshev}
\affiliation{ Irkutsk State University, Irkutsk, 664003 Russia}

\author{M.V.~Lisitsin}
\affiliation{National Research Nuclear University MEPHI, Moscow, Russia, 115409}

\author{S.V.~Lovtsov}
\affiliation{ Irkutsk State University, Irkutsk, 664003 Russia}

\author{R.R.~Mirgazov}
\affiliation{ Irkutsk State University, Irkutsk, 664003 Russia}

\author{D.V.~Naumov}
\affiliation{ Joint Institute for Nuclear Research, Dubna, 141980  Russia}

\author{A.S.~Nikolaev}
\affiliation{ Skobeltsyn Research Institute of Nuclear Physics, Moscow State University, Moscow, 119991 Russia}

\author{I.A.~Perevalova}
\affiliation{ Irkutsk State University, Irkutsk, 664003 Russia}

\author{A.A.~Petrukhin}
\affiliation{National Research Nuclear University MEPHI, Moscow, Russia, 115409}

\author{D.P.~Petukhov}
\affiliation{ Institute for Nuclear Research of the Russian Academy of Sciences, 
Moscow, 117312 Russia}

\author{E.N.~Pliskovsky}
\affiliation{ Joint Institute for Nuclear Research, Dubna, 141980  Russia}

\author{M.I.~Rozanov}
\affiliation{ St. Petersburg State Marine Technical University, St. Petersburg, 190008 Russia}

\author{E.V.~Ryabov}
\affiliation{ Irkutsk State University, Irkutsk, 664003 Russia}

\author{G.B.~Safronov}
\affiliation{ Institute for Nuclear Research of the Russian Academy of Sciences, 
Moscow, 117312 Russia}

\author{B.A.~Shaybonov}
\affiliation{ Joint Institute for Nuclear Research, Dubna, 141980  Russia}

\author{V.Y.~Shishkin}
\affiliation{Skobeltsyn Institute of Nuclear Physics MSU, Moscow, Russia, 119991}

\author{E.V.~Shirokov}
\affiliation{ Skobeltsyn Research Institute of Nuclear Physics, Moscow State University, Moscow, 119991 Russia}

\author{F.~\v{S}imkovic}
\affiliation{ Comenius University, Bratislava, 81499 Slovakia}
\affiliation{Czech Technical University, Institute of Experimental and Applied Physics, CZ-11000 Prague, Czech Republic}

\author{A.E. Sirenko}
\affiliation{ Joint Institute for Nuclear Research, Dubna, 141980  Russia}

\author{A.V.~Skurikhin}
\affiliation{ Skobeltsyn Research Institute of Nuclear Physics, Moscow State University, Moscow, 119991 Russia}

\author{A.G.~Solovjev}
\affiliation{ Joint Institute for Nuclear Research, Dubna, 141980  Russia}

\author{M.N.~Sorokovikov}
\affiliation{ Joint Institute for Nuclear Research, Dubna, 141980  Russia}

\author{I.~\v{S}tekl}
\affiliation{Czech Technical University, Institute of Experimental and Applied Physics, CZ-11000 Prague, Czech Republic}

\author{A.P.~Stromakov}
\affiliation{ Institute for Nuclear Research of the Russian Academy of Sciences, 
Moscow, 117312 Russia}

\author{O.V.~Suvorova}
\email{osouvorova@gmail.com}
   \affiliation{ Institute for Nuclear Research of the Russian Academy of Sciences, 
Moscow, 117312 Russia}

\author{V.A.~Tabolenko}
\affiliation{ Irkutsk State University, Irkutsk, 664003 Russia}

\author{V.I.~Tretjak}
\affiliation{ Joint Institute for Nuclear Research, Dubna, 141980  Russia}

\author{B.B.~Ulzutuev}
\affiliation{ Joint Institute for Nuclear Research, Dubna, 141980  Russia}

\author{Y.V.~Yablokova}
\affiliation{ Joint Institute for Nuclear Research, Dubna, 141980  Russia}

\author{D.N.~Zaborov}
\affiliation{ Institute for Nuclear Research of the Russian Academy of Sciences, 
Moscow, 117312 Russia} 

\author{S.I.~Zavjalov}
\affiliation{ Joint Institute for Nuclear Research, Dubna, 141980  Russia}

\author{D.Y.~Zvezdov}
\affiliation{ Joint Institute for Nuclear Research, Dubna, 141980  Russia}

\collaboration{Baikal-GVD Collaboration}\noaffiliation

\date{\today}

\begin{abstract}
We present an updated measurement of the diffuse astrophysical neutrino flux using Baikal-GVD cascade data collected between April 2018 to March 2024. In this period, the detector grew from 15\% to 55\% of its baseline cubic kilometer configuration. The diffuse astrophysical neutrino flux is detected with a statistical significance of 5.1 $\sigma$.
Assuming a single power law model of the astrophysical neutrino flux with 
flavor equipartition,
the following best-fit parameter values are found: the spectral index $\gamma_{astro}$ = 2.64$^{+0.09}_{-0.11}$ and the flux normalization $\phi_{astro}$ = 4.42$^{+2.31}_{-1.29}\times10^{-18} \text{GeV}^{-1}\text{cm}^{-2}\text{s}^{-1}\text{sr}^{-1}$ per one flavor at 100~\text{TeV}. These results are broadly 
consistent with IceCube measurements.
\end{abstract}

\maketitle

\section{Introduction} \label{s:intro}
It is widely accepted that astrophysical objects, such as, e.g., supernova remnants and active galactic nuclei, can be sources of high-energy neutrino fluxes (see, e.g., ~\cite{LM2000,Becker2008}).
The global flux from a large number of isotropically  distributed sources would be an approximately isotropic diffuse flux. Indeed, such a diffuse flux of astrophysical neutrinos has been discovered by IceCube in 2013~\cite{IC:2013}.

Theoretical models of high energy neutrino production typically involve the acceleration of charged cosmic rays with subsequent $pp$ or $p\gamma$ interactions, which consequently lead to charged pions, their further decays and neutrino production~\cite{Pro:1983,Kaz:1986,Sikora:1987,Stecker:1992,Man:1992}.
In this case, the neutrino flux emitted by a source consists of neutrinos of three flavours in a proportion $\nu_e : \nu_{\mu} : \nu_{\tau} \approx 1 : 2 : 0$. Due to the effect of neutrino oscillations the flavour ratio evolves with neutrino propagation. Since the oscillation length is considerably smaller than the characteristic distances from the source to the detector, the flavour ratio becomes $\nu_e : \nu_{\mu} : \nu_{\tau} \approx 1 : 1 : 1$~\cite{LP:1995,Athar:2006}.

Baikal-GVD is a large volume neutrino telescope of the second generation of Cherenkov detectors in natural reservoirs, similar to IceCube~\cite{IC-HL:2023} and KM3NeT~\cite{km3net-HL:2017}. 
A cubic-kilometer size of the Baikal-GVD detector is planned to be achieved in the next three years. The telescope observes neutrinos in deep water of Lake Baikal by capturing the Cherenkov light emitted by secondary charged particles produced in neutrino interactions with matter in the detector's vicinity. Events which do not involve a high-energy muon are classified as "cascade events". These include neutral-current neutrino interactions of all neutrino flavours and charged-current  interactions of electron and tau neutrinos. Cascade events tend to be contained within the detector volume and allowing for a "calorimetric" measurement of the deposited energy with an accuracy of 10--20\% in dependence on optical characterizations of water. Together with the relatively low atmospheric neutrino background, this makes cascade events especially well suited for spectral measurements of diffuse neutrino fluxes, both for the case of upward-going events and of very high energy events in all-sky observations.

IceCube studied the diffuse flux of high-energy astrophysical neutrinos by employing various  datasets~\cite{IC:2021,IC:2111,IC:2020,IC:2018}. A diffuse emission of neutrinos from the Milky Way has been identified by IceCube in 2024~\cite{IC:2023}. 

Baikal-GVD's modular design allows for data taking during the construction phase. In 2022, Baikal-GVD reported its first observation of the diffuse astrophysical neutrino flux -- with a 3 $\sigma$ significance -- using data collected between April 2018 and March 2022 (4 years)~\cite{GVD:2023}. The presence of a large Galactic component in the high-energy astrophysical neutrino flux was also found in the analysis of high-energy cascade data observed by Baikal-GVD during 6 yr of operation~\cite{GVD-2025}.

Here we present an updated and improved analysis of the diffuse astrophysical neutrino flux which incorporates two more years of data (up to March 2024). For the first time with Baikal-GVD, a statistical significance greater than 5 sigma is reached, allowing for a more accurate characterization of the observed diffuse flux compared to our previous measurements.

This paper is organized as follows: in Section II the Baikal-GVD telescope is briefly introduced; Section III gives an overview of the data sample and data analysis procedure; analysis results are reported in Section IV; finally, Section V concludes the paper.

\section{BAIKAL-GVD NEUTRINO EXPERIMENT} \label{s:GVD}

Baikal-GVD is currently the largest neutrino telescope operating in the Northern Hemisphere~\cite{Baikal-YADFIZ:2022}, with a detection volume of about 0.7 km$^3$ achieved in April 2025. The telescope is located in the Southern part of Lake Baikal ($51^\circ46^\prime$ N, $104^\circ24^\prime$ E) at about 4 km from the shore and at the depth of deployment of $\sim$1366 m. The detector is formed by sub-arrays (clusters) of  optical modules (OM) instrumented with 10-inch high-quantum-efficiency PMT HAMAMATSU36 R7081-100 and various sensors. Each cluster comprises 288 OM distributed on 8 vertical strings and hosted between depths of 750 and 1275 meters. The connection of each cluster to the shore station by its own electro-optical cable provides both independent detection of events by single cluster and by multi-clusters in their time-synchronized operations. 

\begin{figure}
    \centering
 \includegraphics[width=\linewidth]{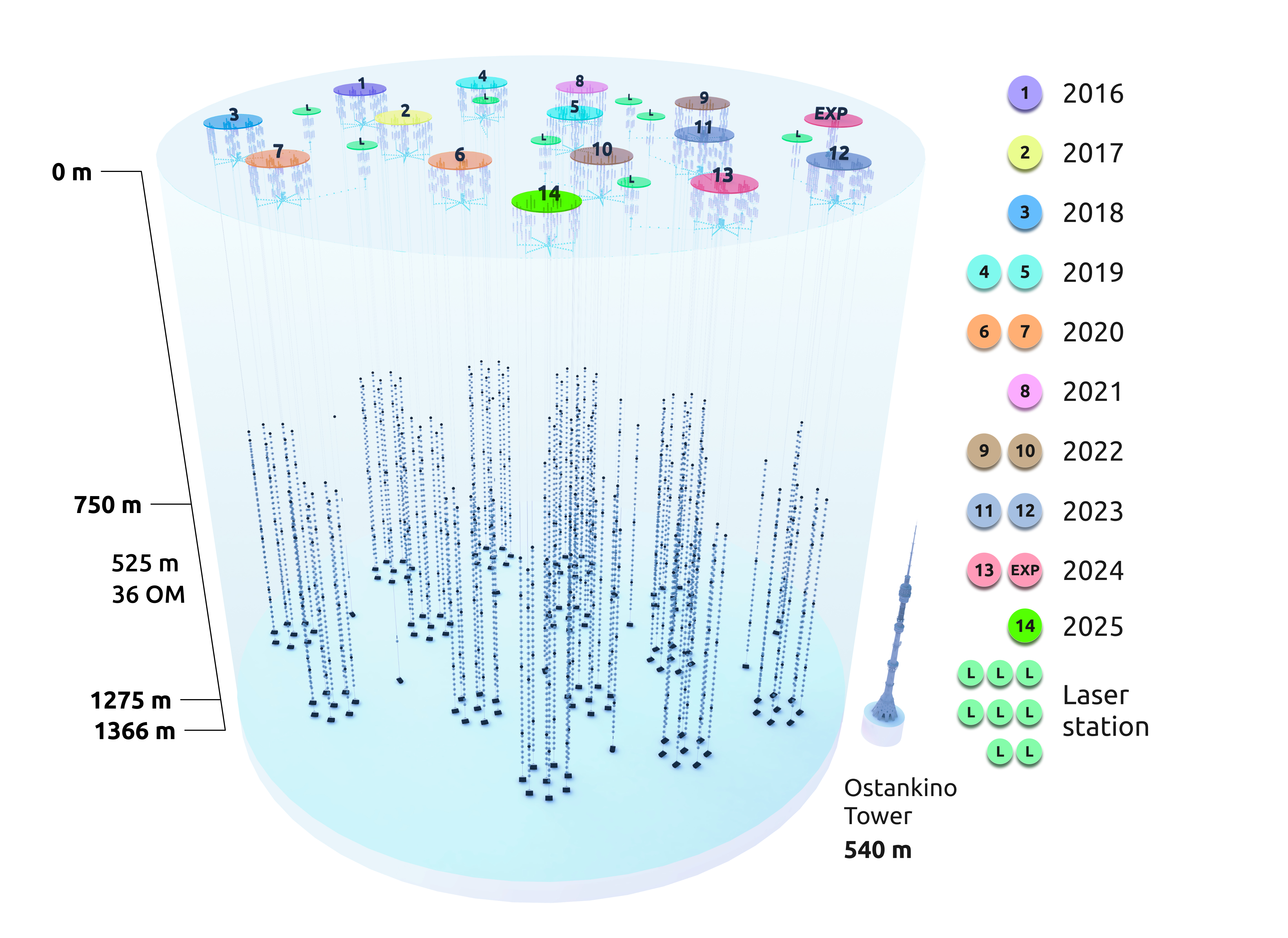}
    \caption
    {Schematic view of the Baikal-GVD configuration achieved by April 2025. The present work uses data from clusters 1--12.
    }
    \label{f:fig1}
\end{figure}

The first full-scale Baikal-GVD cluster was deployed in April 2016. After winter expedition in 2025 the telescope incorporates 14 clusters (see Fig.\ref{f:fig1}) including in total 117 strings carrying 4212 OMs. The design and basic characteristics of the telescope data acquisition system are described elsewhere~\cite{Baikal-YADFIZ:2022,Baikal-MUON}.

\section{DATA ANALYSIS} \label{s:DATAANALYS}
The search for high-energy astrophysical neutrinos is based on
the selection and reconstruction of high-energy showers generated 
in the telescope detection volume. 
In this analysis we used the event selection and shower reconstruction procedures similar to those used in our previous analysis~\cite{GVD:2023}. We select events with a multiplicity of triggered OMs, $N_\mathrm{hit}>7$ at three or more strings, and requiring that hits have charges $Q > 1.5$~p.e. and satisfy the causality condition~\cite{Baikal-MUON}.

The shower vertex coordinates $\vec{r}_\mathrm{sh}$ are reconstructed  by minimization of 
$\chi^2_\mathrm{t}$ function using the time information from the selected hits~\cite{Baikal-ICRC17}.
The shower energy ($E_\mathrm{sh}$) and direction ($\theta, \phi$) are reconstructed by applying the maximum-likelihood method with the use of the reconstructed shower coordinates.
Poorly reconstructed events are rejected by applying cuts on quality parameters, including the values of $\chi^2_\mathrm{t}$ and maximum-likelihood function and OMs hit multiplicity $N_\mathrm{hit}$~\cite{NT200-HE2}.
The precision of energy reconstruction is typically varied in the range 10--20\%. The precision of reconstruction of the shower direction is 2$^\circ$--4$^\circ$ (median value)~\cite{Baikal-JETP:2022}.
The cosmic ray background was simulated with CORSIKA 7.74~\cite{CORSIKA} using the proton spectrum proposed in~\cite{Gaisser} and the SIBYLL 2.3c interaction model~\cite{SIBYLL}. 
The propagation of muons in water was simulated based on the MUM program~\cite{MUM}. 
The passage of neutrinos through the Earth and the interaction in the sensitive volume of the telescope
were simulated using the neutrino cross sections from~\cite{Lai,Gandhi}, the $\tau$ lepton decay cross sections from~\cite{Lipari}, and the model of the Earth profile from~\cite{Earth}. The telescope response to the Cherenkov radiation of showers from neutrino interactions was simulated accounting for the shower development in water, as well as light absorption, scattering and light velocity dispersion in water.

Astrophysical neutrino event selection efficiencies were tested using a flux
presented by Baikal-GVD in~\cite{GVD:2023}.
The conventional atmospheric neutrino flux from pion and kaon decays was modeled according to~\cite{Volkova}. Atmospheric prompt neutrino were simulated according to the BERSS model~\cite{BERSS:2015}.

\section{RESULTS AND DISCUSSION} \label{s:discussion}
We use Baikal-GVD data collected between April 2018 and March 2024 for the search for astrophysical neutrinos.  
A season of data taken covers temporal period from April to next March. The telescope operated with 3 clusters in 2018–2019 season, 5 clusters in 2019– 2020 season, 7 in 2020–2021 season, 8 clusters in 2021–2022, and 10 clusters in April 2022–early 2023. From April 2023 to March 2024, the configuration included 11 complete clusters and one incomplete cluster.
In this study, we report on results of data analysis for individual clusters as independent setups. A sample of 5.49$\times$10$^{10}$ events was collected by the basic trigger of the telescope. After applying noise hit suppression procedures, cascade reconstruction and cuts on reconstruction quality parameters, a sample of 12077 cascades with reconstructed energy $E_\mathrm{sh}>15$~TeV and OM hit multiplicity $N_\mathrm{hit}>$11 was selected. 
It was shown in~\cite{GVD:2023} that restricting the analysis to upward-going directions allows for effective suppression of the atmospheric muon background, thus improving the neutrino sample purity and enabling the extension of the analysis towards lower energies. Cascade-like events with reconstructed energy $E_\mathrm{sh}>$15~TeV, OM hit multiplicity $N_\mathrm{hit}>$11 and reconstructed zenith angle $\cos\theta <$ -0.25 were selected as astrophysical neutrino candidates. A total of 25 events have been selected in the April 2018--March 2024 data sample.  The fraction of background events associated with atmospheric muons and neutrinos in the sample selected using these cuts is expected at a level of 20\%. Finally, after applying additional cuts which suppress events from atmospheric muons and atmospheric muon neutrinos~\cite{Baikal-ICRC21cas}, 18 events have been selected, while 2.8$\pm$1 atmospheric background events are expected (1.9 from atmospheric conventional and prompt neutrinos and 0.9 events from mis-reconstructed atmospheric muons). 
The effect of the uncertainty of the detector response to signal and background is 
evaluated by varying input parameters in the Monte Carlo simulations. The uncertainty of the light absorption length is about $\pm$10\%. Such variations lead to a change in the detection efficiency of high-energy cascades by about 18\%-20\% and shift the energy scale in the logarithm of the cascade energy by about $\pm$0.05. At the same time, the uncertainty of the reconstruction of the cascade direction weakly depends on the changes in the light absorption length. The optical module sensitivity varies within $\pm$10\%. Also a $\pm$15\% uncertainty on the normalization of the conventional atmospheric neutrino component is considered~\cite{Volkova}.  It is shown below that the total statistical and systematic uncertainty of the best fit astrophysical flux normalization is about 30\%-50\%. Given an upper limit on the prompt atmospheric neutrino flux normalization about 5.0$\times\Phi_\textit{BERSS}$, as it was assumed in IceCube’s 6 yr cascade analysis~\cite{IC:2020}, an uncertainty of the prompt neutrino flux normalization would cause the uncertainty of the astrophysical best fit flux normalization less than 4\%. For this reason, this uncertainty was not included in this analysis. The uncertainties coming from independent sources are added in quadrature in the overall estimation. Taking into account the systematic effects according to the method of~\cite{Conrad:2003}, the significance of the excess was estimated to be 5.1$\sigma$ with the chance probability 2.1$\cdot10^{-7}$.
 The parameters of the 18 upward-going cascades are shown in Tab. I. The median value of the error in the reconstruction of the cascade direction varies from 1.6$^\circ$ to 5.4$^\circ$.
\begin{figure}[ht]
    \centering
  \includegraphics[width=0.9\linewidth]{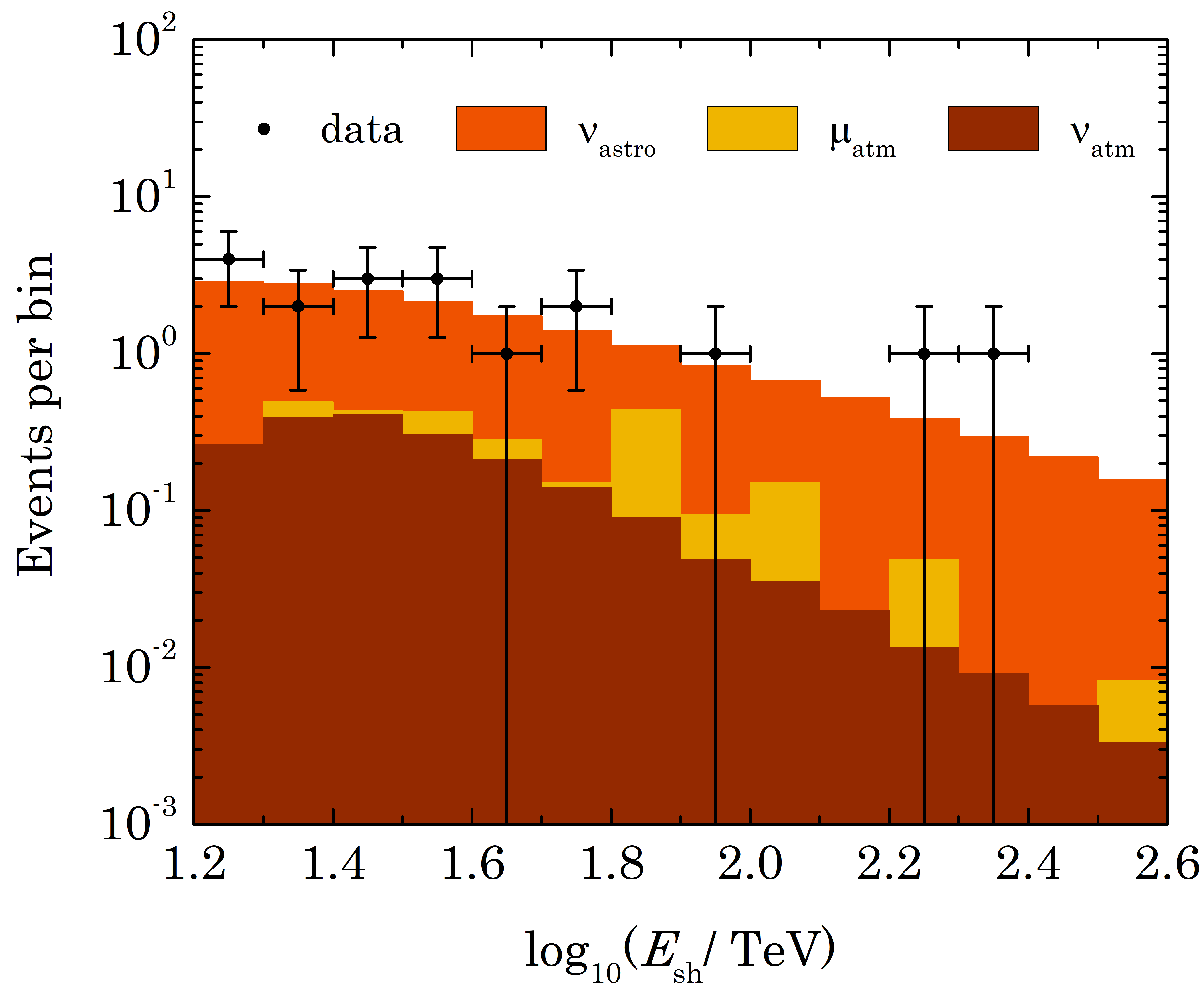}
  \includegraphics[width=0.9\linewidth]{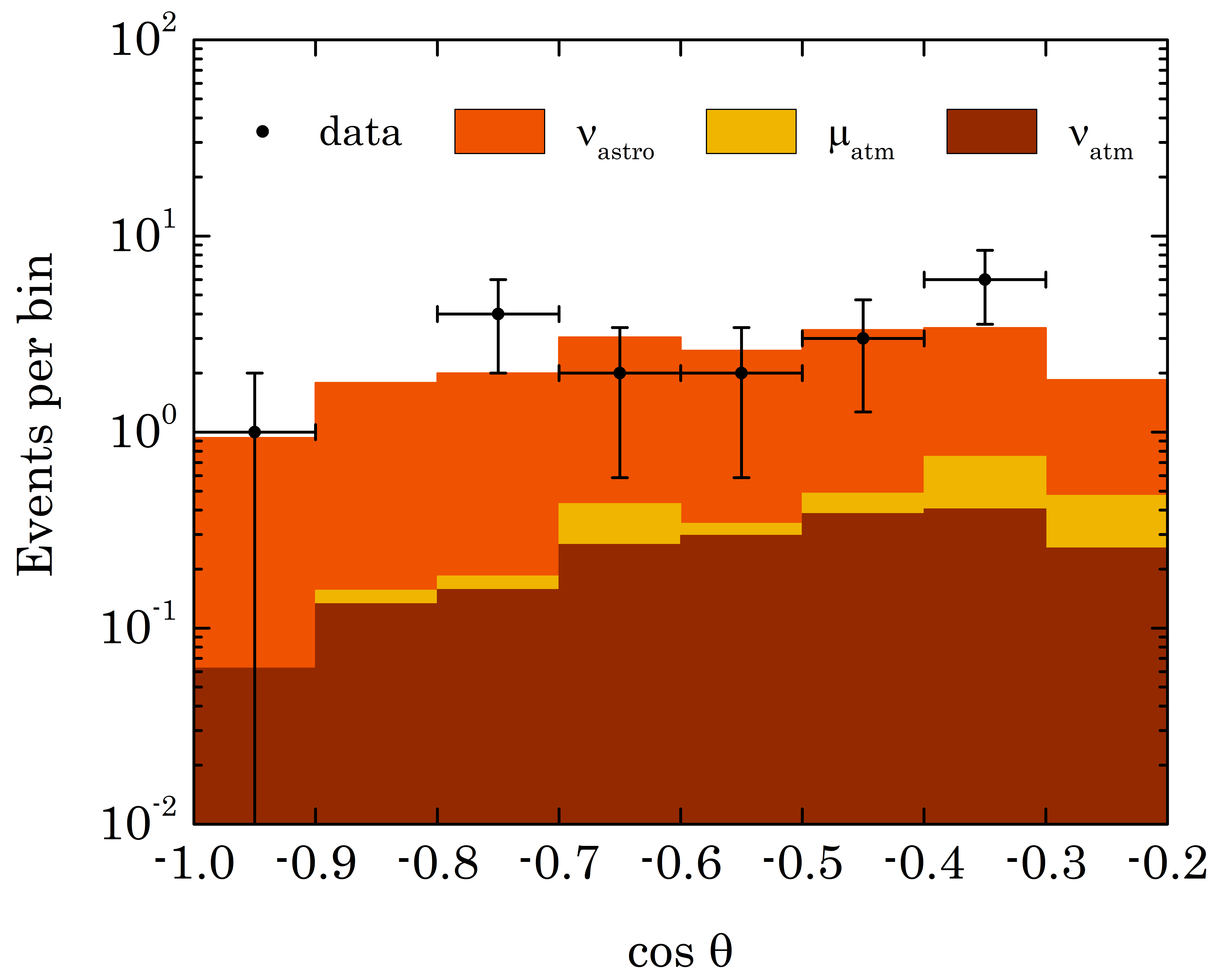}

    \caption{Reconstructed cascade energy (top panel) and zenith (bottom panel) distributions obtained in the upward-going cascade analysis. Black points are data, with statistical uncertainties. The stacked colored bands show the expected contribution from background atmospheric neutrinos (brown) and atmospheric muons (yellow), as well as from the best fit astrophysical neutrino flux obtained in this work (dark orange).
      }
    \label{f:fig4}
\end{figure}
The measured 18 events and the expected number of background events have been analyzed to characterize the diffuse astrophysical neutrino flux. We parameterize the isotropic diffuse astrophysical neutrino flux $\Phi^{\nu+\bar{\nu}}_{astro}$ in the single power law model assuming equal numbers of neutrinos and antineutrinos and equal neutrino flavors ratio at Earth. The model is characterized by spectral index $\gamma_{astro}$ and normalization 
$\phi_{astro}$ of the one-flavour neutrino flux in units of $10^{-{18}}\text{GeV}^{-{1}}\text{cm}^{-{2}}\text{s}^{-{1}} \text{sr}^{-{1}}$: 
\begin{equation}
\Phi^{\nu+\bar{\nu}}_{astro} = \phi_{astro} \left(\frac{E_{\nu}}{E_0}\right)^{-{\gamma_{astro}}},
\end{equation}
where $E_0$ = 100~TeV. By the same way as in our previous analysis~\cite{GVD:2023} the best fit parameters for the observed data are determined by a binned likelihood approach. In this procedure, the data sample is binned in reconstructed shower energy.
The observed count $n_i$ in each bin $i$ is compared to a model that predicts the mean count rate $\lambda_i$ in each bin using a Poisson likelihood function.
The expected rates $\lambda_i$ are composed by astrophysical neutrinos and background events of atmospheric muons and atmospheric neutrinos. The MC simulated templates of the cosmic signal and of the atmospheric backgrounds with different water parameters and OM efficiency were used for $\lambda_i$ estimation. Accordingly, these templates include effects of change of detection efficiency and energy scale shift. Following the Poisson likelihood function (3) and the compiled test statistic (TS) as in our previous study~\cite{GVD:2023}, we fitted the observed event counts by the Monte Carlo model predictions.
The systematic uncertainties discussed above were incorporated in the test statistic as ($k$) nuisance parameters in form of Gaussian distributions of prior $\Psi_k$ and  width deviation $\sigma(\Psi_k)$ from central value $\Psi^0_k$. A maximum-likelihood method is applied to find the best-fit values of $\gamma_{astro}$ and $\phi_{astro}$ by varying these parameters until TS is minimized. We find the best-fit parameters as follows: the spectral index $\gamma_{astro}$ = 2.64$^{+0.09}_{-{0.11}}$ and the flux normalization for each neutrino flavour at $E_0$ = 100~TeV $\phi_{astro}$ = 4.42$^{+2.31}_{-1.29}\times10^{-18} \text{GeV}^{-1}\text{cm}^{-2}\text{s}^{-1}\text{sr}^{-1}$.
\begin{figure}[ht]
    \centering
\includegraphics[width=0.9\linewidth]{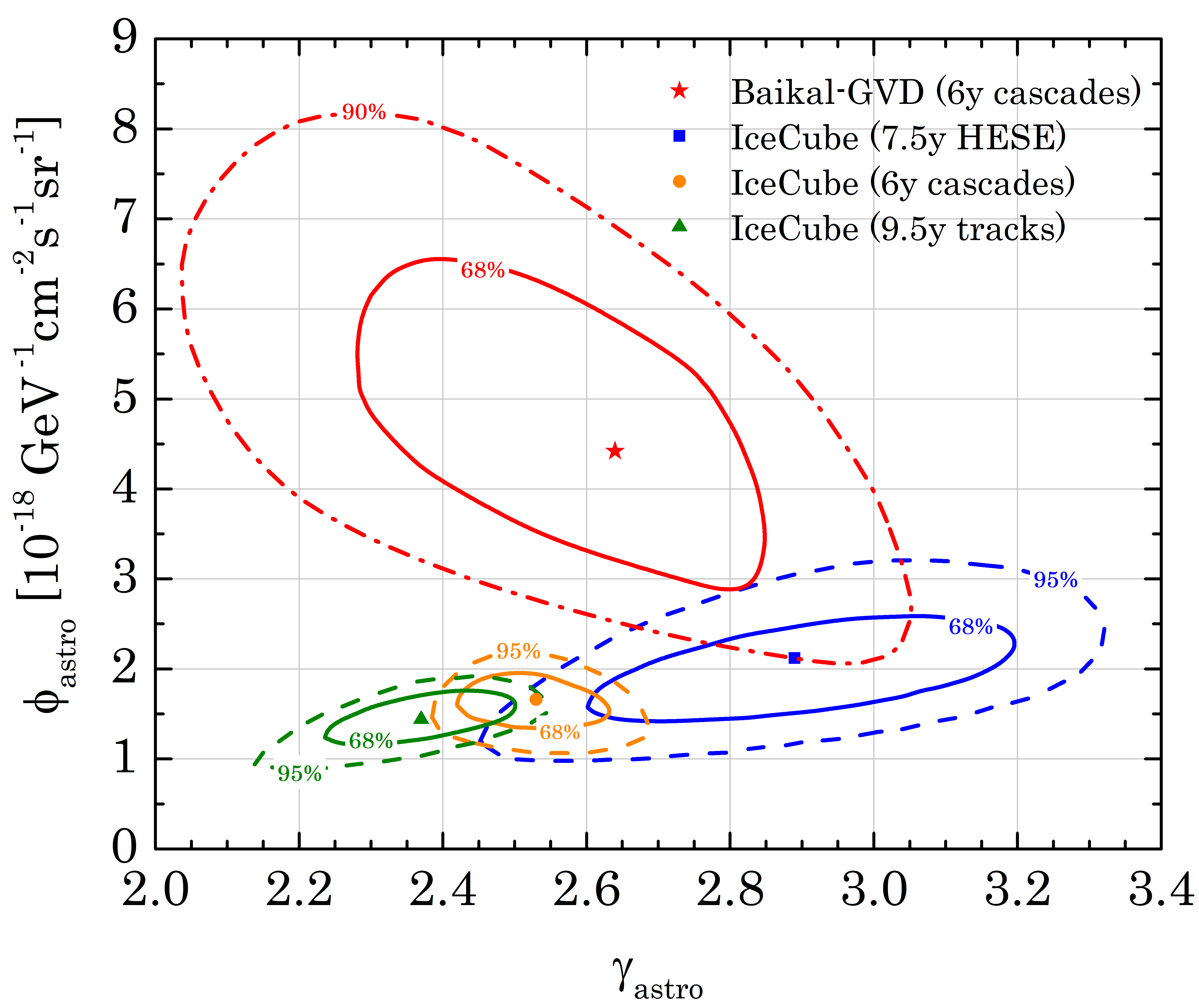}
 \caption{The best fit parameters (red star) and the contours at 68\% confidence region (red solid curve) and 90\% one (red dot-dashed) for the single power law hypothesis 
 obtained in the upward-going cascade analysis of the Baikal-GVD data. Other best fits and the confidence level contours of the 68\% (solid) and 95\% (dashed) are shown for studies based on high-energy starting events (blue)~\cite{IC:2021}, track-like events (green)~\cite{IC:2111} 
 and cascade-like events (light orange)~\cite{IC:2020} by IceCube.
 }
    \label{f:fig6x}
\end{figure}
\begin{figure}[ht]
    \centering
\includegraphics[width=0.85\linewidth]{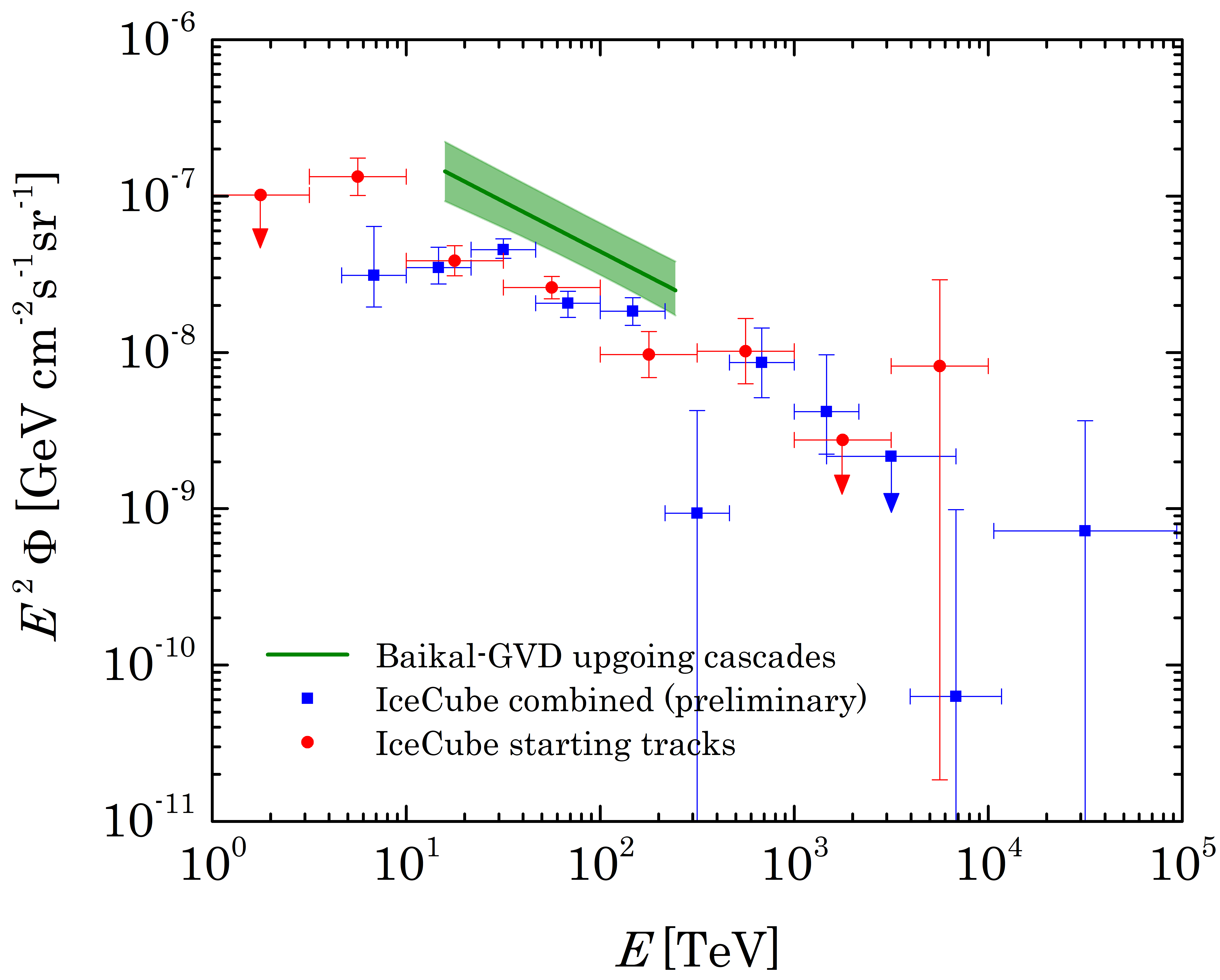}
 \caption{Measurements of the diffuse astrophysical neutrino flux: the green line shows the fitted spectrum for this analysis of the Baikal-GVD upward-going cascade sample and the uncertainties are shown by the shaded region. The red points show differential data points of the latest IceCube analysis of starting tracks sample~\cite{IceCube:2024fxo}, while the blue points represent the IceCube preliminary result of their combined analysis~\cite{Naab:2023xcz}. 
 }
    \label{f:fig7x}
\end{figure}
\begin{table*}[ht]
\caption{Parameters of 18 under horizon cascade events: date of observation as Modified Julian Date, reconstructed energy, zenith angle, Galactic longitude and latitude, right ascension and declination, 50$\%$ and 90$\%$-containment angular uncertainty region, distance between shower vertex and central string of cluster. The event name (left column) encodes the event detection date in the format yymmdd.}
\begin{ruledtabular}
\begin{tabular}{ccccrrrrccc}
 Event name & MJD & E$_{sh}$ & $\theta$ & $l$ & $b$ & RA & Dec & 50$\%$ unc. & 90$\%$ unc. & $\rho$\\
 &  & TeV & deg. & deg. & deg. & deg. & deg. & deg. & deg. & meter \\
\hline
GVD190523CA &58626.44462963 & 91.0 & 109.0  &  200.4  &  $-$58.4& 45.1  & $-$16.7  & 2.2 & 4.5   & 49 \\
GVD201112CA &59165.01353009 & 24.5 & 136.1  &  305.0  &  $-$15.1& 202.2 & $-$77.8  & 5.4 & 11.8  &  66 \\
GVD210418CA &59322.94855324 & 224  & 115.5  &  196.8  &  $-$14.6 & 82.4  & 7.1     & 3.0 & 5.8   &  70\\
GVD210506CA &59340.34252315 & 21.9 & 114.2  &  5.9    &  46.7   & 230.6 & 3.1      & 2.8 & 6.6   &  30\\
GVD220121CA &59600.45934028 & 30.9 & 110.5  &  241.3  &  10.4   & 126.2 & $-$19.5  & 3.4 & 7.1   &  49\\
GVD220406CA &59675.72173611 & 46.6 & 117.7  &  223.1  &  $-$21.6& 87.4  & $-$18.3  & 3.5 & 9.1 &  81 \\
GVD220625CA &59755.81578704 &134.9 & 108.7  &  292.9  &     63.9&188.9  &     1.3  & 2.3 & 4.9   &  69 \\
GVD220711CA &59771.24636574 & 16.6 & 133.0  &   37.4  &  $-$71.6&352.7  & $-$24.4  & 3.8 & 8.1   &  49 \\
GVD220805CA &59797.05288194 & 15.1 & 109.3  &   33.4  &     41.6&244.7  &    18.0  & 2.5 & 5.1   &  77 \\
GVD220814CA &59805.36122685 & 53.7 & 126.9  &  221.2  &  $-$34.3& 74.1  & $-$21.3  & 1.6 & 3.2   &  64 \\
GVD221124CA &59908.16281250 & 29.5 & 125.5  &  177.0  &  $-$45.0& 47.8  &     2.7  & 3.2 & 6.8   & 104 \\
GVD221211CA &59925.44519676 & 17.0 & 107.5  &  212.3  &     32.4&132.2  &    14.6  & 2.6 & 7.0   &  68 \\
GVD230817CA &60173.50042824 & 25.7 & 129.8  &  151.8  &  $-$86.4& 14.8  &  $-$24.0 & 2.8 & 6.8   &  84 \\
GVD231014CA &60231.85099537 & 34.7 & 141.0  &  318.1  &     29.8& 208.2 &  $-$31.2 & 2.4 & 4.5   &  65 \\
GVD230814CA &60170.35375000 & 53.7 & 107.6  &  255.6  &   $-$0.5& 125.1 &  $-$37.4 & 3.1 & 6.0   &  25 \\
GVD230529CA &60094.35293981 & 20.0 & 168.5  &  333.3  &  $-$31.9& 300.5 &  $-$63.2 & 3.7 & 8.6   &  67 \\
GVD240201CA &60342.30295139 & 32.4 & 142.2  &  321.8  &     24.1& 214.1 &  $-$35.7 & 2.9 & 6.3   &  77 \\
GVD230820CA &60176.39195602 & 15.5 & 126.2  &  287.8  &   $-$7.7& 153.6 &  $-$65.8 & 3.2 & 7.1   &  49 \\
\end{tabular}
\end{ruledtabular}
\label{tab:hor_events}
\end{table*}

The energy and zenith distributions of the 18 events are shown in Fig.\ref{f:fig4} together with the distributions obtained by Monte Carlo simulation. The atmospheric background histograms and best fit astrophysical flux histogram are stacked (filled colors). The best-fit parameters as well as 68\% C.L. and 90\% C.L. contours for this cascade analysis together with the results from other neutrino telescopes~\cite{IC:2021,IC:2111,IC:2020,IC:2018}
are shown in Fig~\ref{f:fig6x}. The Baikal-GVD upward-going neutrino (cascades) measurements are generally consistent with the IceCube measurements~\cite{IC:2021} as also seen in Fig~\ref{f:fig7x}, which shows the Baikal-GVD results of this work and the latest fits of the IceCube sample of starting tracks~\cite{IceCube:2024fxo} and their combined analysis for different data samples~\cite{Naab:2023xcz}. 
However, the Baikal-GVD data favour a somewhat higher flux than recent IceCube measurements. Upcoming data and further analyses will be essential to determine whether this difference persists and to refine the diffuse-flux estimate.

\section{Conclusion} \label{s:summary}
We presented the measurements of astrophysical neutrino flux using a sample of upward moving cascade events with energy $E_\mathrm{sh}>$15 TeV collected by the Northern Hemisphere neutrino telescope Baikal-GVD in April 2018--March 2024.
A total of 18 events have been selected as astrophysical neutrino candidates, while 2.8 atmospheric background events are expected. The significance of the excess over the expected number of atmospheric background events was estimated as 5.1$\sigma$. 
The cascade energy distribution has been fitted with a single power-law model for the astrophysical flux and MC-based templates for the atmospheric backgrounds, taking into account major sources of detector and water related systematic uncertainties.
The measured values of the spectral index of astrophysical neutrinos,
$\gamma_{astro}$ = 2.64$^{+0.09}_{-{0.11}}$, and the per-flavor flux normalization for each neutrino flavor at $E_0$ = 100 TeV $\phi_{astro}$ = 4.42$^{+2.31}_{-1.29}\times10^{-18} \text{GeV}^{-1}\text{cm}^{-2}\text{s}^{-1}\text{sr}^{-1}$  are 
consistent with the earlier version of the Baikal-GVD analysis~\cite{GVD:2023}. This constitutes the first independent observation of the astrophysical neutrino flux at a significance level exceeding 5$\sigma$, achieved by Baikal-GVD based solely on upward-moving cascade events.

\begin{acknowledgments}
\begin{itemize}
    \item We acknowledge Sergey Troitsky for fruitful discussion and partnership. This work is supported in the framework of the State project “Science” by the Ministry of Science and Higher Education of the Russian Federation under the contract 075-15-2024-541.
\end{itemize}
\end{acknowledgments}



\bibliography{Baikal-GVD_DiffuseFlx_2025} 

\end{document}